\documentclass[aps,twocolumn,showpacs]{revtex4}
\usepackage{amssymb,graphicx}
\usepackage{amsmath,bbm}
\begin{document}
%%%%%%%%%%%%%%%%%%%%%%%%%%%%%%%%%%%%%%%%%%%%%%%%%%%%%%%%%%%%%%%%
\title{Optimal quantum repeaters for qubits and qudits}
\author{Marco G. Genoni and Matteo G. A. Paris} 
\email{matteo.paris@fisica.unimi.it}
\affiliation{Dipartimento di Fisica and INFM, Universit\`a 
degli studi di Milano, Italia.}
%%%%%%%%%%%%%%%%%%%%%%%%%%%%%%%%%%%%%%%%%%%%%%%%%%%%%%%%%%%%%%%%
\date{\today}
\begin{abstract} 
A class of optimal quantum repeaters for qubits is suggested.
The schemes are {\em minimal}, {\em i.e.} involve a single 
additional {\em probe} qubit, and {\em optimal}, {\em i.e.} 
provide the maximum information adding the minimum amount of noise.  
Information gain and state disturbance are quantified by fidelities 
which, for our schemes, saturate the ultimate bound imposed by 
quantum mechanics for randomly distributed signals. 
Special classes of signals are also investigated, in order 
to improve the information-disturbance trade-off. Extension 
to higher dimensional signals (qudits) is straightforward.
\end{abstract}
\maketitle
%%%%%%%%%%%%%%%%%%%%%%%%%%%%%%%%%%%%%%%%%%%%%%%%%%%%%%%%%%%%%%%%
%\special{!userdict begin /bop-hook{gsave 35 20 
%translate 90 rotate /Times-Roman findfont 30 scalefont setfont
%60 0 moveto 0.4 setgray (From Matteo to Radim, draft version, 
%7 Jan 2005) show grestore 
%}def end}
%%%%%%%%%%%%%%%%%%%%%%%%%%%%%%%%%%%%%%%%%%%%%%%%%%%%%%%%%%%%%
\section{Introduction}\label{s:intro}
In a multiuser transmission line each user should decode the
transmitted symbol and leave the carrier for the subsequent user.
What they need is an ideal {\em repeater}, {\em i.e.} a device
that for each shot retrieves the message without altering the
carrier. However, symbols are necessarily encoded in states of
a physical system and therefore the ultimate bound on the
performances as a repeater are posed by quantum mechanics. 
Indeed, a perfect quantum repeater cannot be achieved, {\em i.e} 
quantum information cannot be perfectly copied, neither locally 
\cite{nocl} nor at distance \cite{telecl}. Any measurement
performed to extract information on a quantum state in turn
alters the state itself, {\em i.e.} produces a disturbance.
\par
The trade-off between information gain and quantum state
disturbance can be quantified using fidelities. Let us describe a
generic scheme for indirect measurement as a quantum operation,
{\em i.e.} without referring to any explicit unitary realization.
The operation is described by a set of {\em measurement operators
$\{A_k\}$}, with the condition $\sum_k A^\dag_k A_k= {\mathbbm
I}$. The probability-operator measure (POVM) of the measurement
is given by $\{\Pi_k\equiv A_k^\dag A_k\}$, whereas its action on
the input state is expressed as $\varrho \rightarrow \sum_k
A_k\varrho A_k^\dag$. This means that, if $\varrho$ is the
initial quantum state of the system under investigation, the
probability distribution of the outcomes is given by
$p_k=\hbox{Tr}[\varrho\: \Pi_k]= \hbox{Tr}[\varrho\: A^\dag_k
A_k]$, whereas the conditional output state, after having
detected the outcome $k$, is expressed as $\sigma_k= \:
A_k\varrho A_k^\dag/p_k$, such that the overall quantum state
after the measurement is described by the density matrix
$\sigma=\sum_k p_k \: \sigma_k=\sum_k A_k\varrho A_k^\dag$.  
\par
Suppose now you have a quantum system prepared in a pure state
$|\psi\rangle$. If the outcome $k$ is observed at the output
of the repeater, then the estimated signal state is given by
$|\phi_k\rangle$ (the typical inference rule being 
$k \rightarrow |\phi_k\rangle$ with $|\phi_k\rangle$ given 
by the set of eigenstates of the measured observable), 
whereas the conditional state $|\psi_k\rangle = 1/\sqrt{p_k} 
A_k |\psi\rangle$ is left for the subsequent user. The amount 
of disturbance is quantified by evaluating the
overlap of the conditional state $|\psi_k\rangle$ to 
the initial one $|\psi\rangle$, whereas the
amount of information extracted by the measurement 
corresponds to the overlap of the inferred state $|\phi_k\rangle$ 
to the initial one. 
The corresponding fidelities, for a given input signal 
$|\psi\rangle$, are given by
\begin{eqnarray}
F_\psi &=& \sum_k p_k \frac{|\langle\psi|A_k |\psi\rangle |^2}{p_k} 
= \sum_k |\langle \psi|A_k|\psi\rangle |^2 \label{F_psi}
\\ G_\psi &=& \sum_k p_k |\langle\psi|\phi_k\rangle |^2  
\label{G_psi}\;,
\end{eqnarray}
where we have already performed the average over the 
outcomes. The relevant quantities to assess the
repeater are then given by the average fidelities
\begin{eqnarray}
F = \int_{\mathbbm A} d\psi \: F_\psi \qquad
G = \int_{\mathbbm A} d\psi \: G_\psi
\label{fids}\;,
\end{eqnarray}
which are obtained by averaging $F_\psi$ and $G_\psi$ 
over the possible input states, {\em i.e.} over the 
alphabet ${\mathbbm A}$ of 
transmittable symbols. $F$ will be referred to as the 
transmission fidelity and $G$ as the estimation fidelity.
\par
Let us first consider two extreme cases.
If nothing is done, the signal is preserved and thus
$F=1$. However, at the same time, our estimation has to be 
random and thus $G=1/d$ where $d$ is the dimension of the 
Hilbert space. This corresponds to a {\em blind} quantum 
repeater \cite{szeged} which re-prepares any quantum state 
received at the input, without gaining any information on it.
The opposite case is when the maximum information is 
gained on the signal, {\em i.e.} when the optimal estimation 
strategy  for a single copy is adopted \cite{popescu,acin,bruss}. 
In this case $G=2/(d+1)$, but then the signal after this operation 
cannot provide any more information on the initial state and 
thus $F=2/(d+1)$. 
Between these two extrema there are intermediate cases, 
{\em i.e.} quantum measurements providing only partial 
information while partially preserving the quantum state 
of the signal for subsequent users. 
These schemes, which correspond to feasible quantum repeaters, 
may be also viewed as quantum nondemolition measurements 
\cite{bra}, which have been widely investigated for continuous 
variable systems, and recently received attention also for 
qubits \cite{dec04}.
\par
The fidelities $F$ and $G$ are not independent on each other. 
Assuming that ${\mathbbm A}$ corresponds to the set of 
{\em all} possible quantum states, Banaszek \cite{KB} has 
explicitly evaluated the expressions of fidelities in terms of 
the measurement operators, rewriting Eq. (\ref{fids}) as 
\begin{eqnarray}
F &=& \frac{1}{d(d+1)} \left(d+ \sum_k \left| \hbox{Tr} \left[A_k\right]
\right|^2\right)  
\nonumber  \\
G &=& \frac{1}{d(d+1)} \left(d+ \sum_k \langle \phi_k | \Pi_k | 
\phi_k\rangle \right)
\label{fidse}\;,
\end{eqnarray}
where $|\phi_k\rangle$ is the set of states used to estimate the 
initial signal. Of course, the estimation fidelity is maximized 
choosing $|\phi_k\rangle$ as the eigenvectors of $\Pi_k$ 
corresponding to the maximum eigenvalues.
\par
Using (\ref{fidse}) it is possible to derive the bound that fidelities 
should satisfy according to quantum mechanics. 
For randomly distributed $d$-dimensional signals, {\em i.e.} 
when the alphabet ${\mathbbm A}$ corresponds to the set of 
{\em all} quantum states for a qudit, the information-disturbance 
trade-off reads as follows \cite{KB}
\begin{align}
& (F - F_0)^2 + d^2 ( G - G_0)^2  \nonumber \\ 
&+ 2 (d-2)(F - F_0)( G - G_0)
\leq \frac{d-1}{(d+1)^2} \label{Dbound}\;,
\end{align}
where $F_0=\frac12(d+2)/(d+1)$ and $G_0=\frac12\:3/(d+1)$.
For randomly distributed qubits, {\em i.e.} assuming a two-dimensional 
Hilbert space, and with the alphabet ${\mathbbm A}$ equal to the 
whole Bloch sphere, the bound (\ref{Dbound}) reduces to
\begin{eqnarray}
\left(F - \frac{2}{3}\right)^2 + 4\left( G - \frac{1}{2}\right)^2 
\leq \frac{1}{9} \label{bound}\;.
\end{eqnarray}
From Eq. (\ref{Dbound}) one knows the maximum transmission fidelity 
compatible with a given value of the estimation fidelity or, in other
words, the minimum unavoidable amount of noise that is added to the 
knowledge about a set of signals if one wants to achieve a given 
level of information.
\par
In this paper we suggest a set of explicit unitary realizations 
for the indirect estimation of qubits. Our schemes are {\em minimal}, 
since they involve a single additional probe qubit, and {\em optimal} 
{\em i.e.} the corresponding fidelities saturate the bound 
(\ref{bound}) with the equal sign. The schemes 
can be easily generalized to the case of qudits, yet being minimal
and saturating the bound (\ref{Dbound}). Recently \cite{filip}, 
similar schemes have been suggested, also with the possibility of 
obtaining signal independent fidelities through a twirl operation 
\cite{jmir,isotro}. 
\par
The paper is structured as follows. In Section \ref{s:sz} 
the simplest example of our class of schemes will be described 
in details. Its possible generalizations, involving 
the measurement of a generic spin component, are analyzed in
Section \ref{ss:sm}, whereas generalization to dimension $d$ 
is described in Section \ref{ss:D}. In Section \ref{s:sig} we 
return back to qubits, and consider the transmission of 
signals with quantum states that do not span the entire Hilbert 
space, {\em i.e} of alphabets that are proper subsets of the 
Bloch sphere. It will be shown that discrete alphabets can 
be used to beat the bound (\ref{bound}), whereas continuous 
alphabets different from the whole Bloch sphere lead to inferior 
performances. A general condition on the class of signals to beat 
the bound will be also derived.
Section \ref{s:outro} closes the paper with some concluding 
remarks.
%%%%%%%%%%%%%%%%
\section{Minimal implementation of optimal quantum repeaters}\label{s:sz}
In this section we suggest a measurement scheme to estimate the
state of a generic qubit without its destruction. This scheme is
minimal because it involves a single additional probe qubit, and
optimal because it saturates the bound (\ref{bound}).
\par
\begin{figure}[h]
\begin{center}
\includegraphics[width=0.4\textwidth]{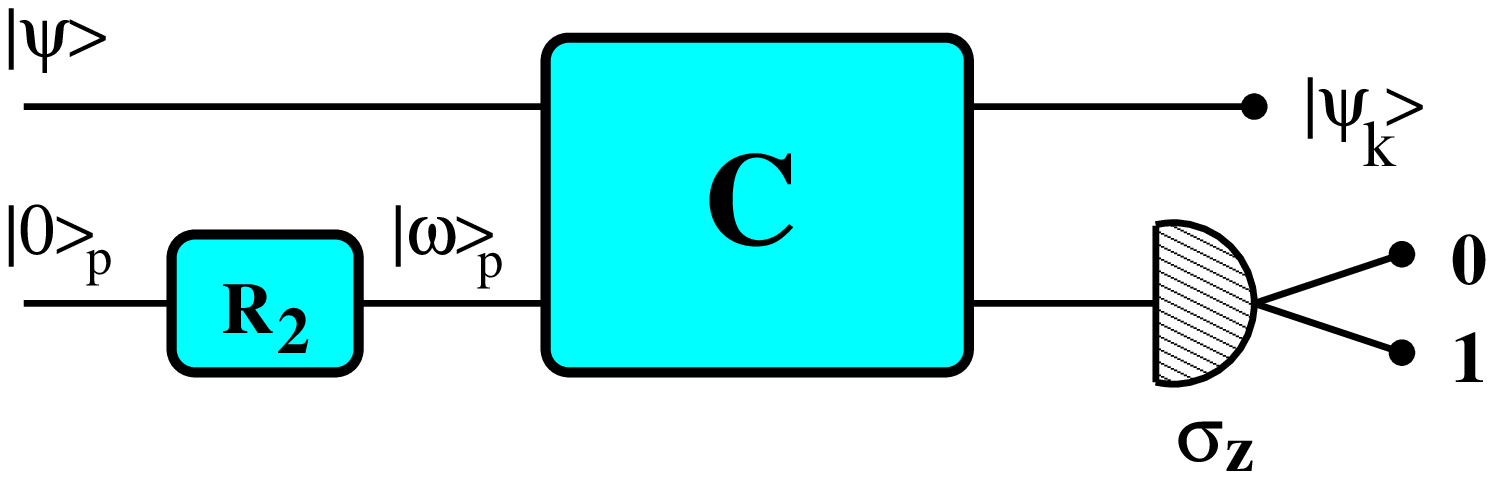}
\end{center}
\caption{Minimal implementation of optimal quantum repeater for qubits.} 
\label{f:sch2}
\end{figure}

The measurement scheme is shown in Fig. \ref{f:sch2}.  
The signal qubit  $$|\psi\rangle = \cos\frac{\theta_1}{2}|0\rangle 
+ e^{i\phi_1}\sin\frac{\theta_1}{2}|1\rangle$$ is coupled with a 
probe qubit $$|\omega\rangle_p = {\mathbf R}_2|0\rangle_p = 
\cos\frac{\theta_2}{2}|0 \rangle_p + e^{i\phi_2}\sin
\frac{\theta_2}{2}|1\rangle_p$$ by a $C_{\rm not}$ gate 
(denoted by ${\mathbf C}$). ${\mathbf R}_i$ denotes a qubit 
rotation, by angles ($\theta_i$,$\phi_i$) with respect to 
the $z$-axis. After the interaction 
the spin component in the $z$-direction is measured on the probe qubit. 
%%%%%%%%%%%%%%%%
%%%%%%%%%%%%%%%%
\par\noindent
According to Eqs. (\ref{F_psi}) and (\ref{G_psi}) the fidelities 
$F_{\psi}$  and $G_{\psi}$ are given by 
\begin{eqnarray}
F_{\psi} &=& p_0|\langle\psi|\psi_0\rangle |^2 + 
p_1|\langle\psi|\psi_1\rangle |^2 
\nonumber \\
G_{\psi} &=& p_0|\langle\psi|0\rangle |^2 + 
p_1|\langle\psi| 1\rangle |^2. \nonumber 
\end{eqnarray}
where $p_k=\langle \psi | \Pi_k | \psi\rangle$, $k=0,1$ are 
the probabilities for the two possible outcomes and 
$$|\psi_k\rangle = \frac{A_k|\psi\rangle}{\sqrt{p_k}}$$ 
the corresponding conditional (pure) states left for the 
subsequent user. 
Moreover, in writing $G_{\psi}$, we assumed the inference rule 
$k \rightarrow |k\rangle$ , with $|k\rangle$  eigenstates of 
the measured observable 
$\sigma_z$.
\par
The measurement operators $A_k$ for our schemes are given by 
\begin{equation}
{A}_k = {}_p\langle k|{\mathbf C}| \omega\rangle_p \label{A_k}\:,
\end{equation}
whereas the POVM can be evaluated as follows 
\begin{equation}
{\Pi}_k = \hbox{Tr}_p\left[ {\mathbf C} \: {\mathbbm I} 
\otimes |\omega\rangle_p{}_p\langle \omega| \: 
{\mathbf C}^{\dag} \: {\mathbbm I} \otimes 
|k\rangle_p{}_p\langle k| \:   \right] = 
{A}_k^{\dag}{A}_k \:, \label{Pi_k}
\end{equation}
where $\hbox{Tr}_p[\ldots]$ denotes partial trace
over the probe' degrees of freedom.
Explicitly, in the standard basis, we have 
$$ A_0 =
\left(
\begin{array}{cc}
\cos\frac{\theta_2}{2} & 0 \\
0 & e^{i\phi_2}\sin\frac{\theta_2}{2} \\
\end{array}
\right)
\quad 
A_1 =
\left(
\begin{array}{cc}
e^{i\phi_2}\sin\frac{\theta_2}{2} & 0 \\
0 & \cos\frac{\theta_2}{2} \\
\end{array}
\right)\:.
$$
The mean fidelities $F$ and $G$ are obtained by averaging 
over all the possible input 
states {\em i.e.} the whole Bloch sphere:
\begin{eqnarray}
F &=& \frac{1}{4\pi}\int_0^{2\pi} d\phi_1 
\int_0^{\pi} d\theta_1 \: \sin\theta_1\: F_{\psi} \nonumber \\
G &=& \frac{1}{4\pi}\int_0^{2\pi} d\phi_1 
\int_0^{\pi} d\theta_1 : \sin\theta_1\: G_{\psi} \nonumber
\end{eqnarray}
According to (\ref{fidse}) this corresponds to 
\begin{eqnarray}
F &=& \frac{1}{6}\left( 2 + \left|\hbox{Tr}\left[{A}_0\right]\right|^2 + 
\left|\hbox{Tr}\left[{A}_1\right]\right|^2 \right) \label{F_2} \\
G &=& \frac{1}{6}\left( 2 + \langle 0|{\Pi}_0| 0 \rangle 
+ \langle 1|{\Pi}_1|
1\rangle \right)\:. \label{G_2} 
\end{eqnarray}
Explicit calculations of formulas (\ref{F_2}) and (\ref{G_2}) leads to 
\begin{eqnarray}
F &=& \frac{1}{6}\left( 2 + 2\left| \cos\frac{\theta}{2} + 
e^{i\phi_2}\sin\frac{\theta_2}{2} \right|^2 \right) 
\nonumber \\ &=& 
\frac{2}{3}\left( 1 + \sin\frac{\theta_2}{2}
\cos\frac{\theta_2}{2}\cos\phi_2 \right) 
\label{F_opt} \\
G &=& \frac{1}{6}\left( 2 + 2\cos^2\frac{\theta_2}{2} \right) 
= \frac{1}{3} \left( 1 + \cos^2\frac{\theta_2}{2} \right)\:.
\label{G_opt}
\end{eqnarray}
Eqs. (\ref{F_opt}) and (\ref{G_opt}) say that any (allowed)
ratio between the two fidelities may be achieved by a suitable
preparation of the probe.
At this point we set $\phi_2 = 0$ and substitute (\ref{G_opt}) 
into (\ref{F_opt}) in order to find the explicit dependence 
$F=F(G)$. We have
\begin{eqnarray}
F = \frac{2}{3}\left( 1 + \sqrt{-9G^2 + 9G - 2} \right) \label{fg}
\end{eqnarray}
The function $F(G)$ in (\ref{fg}) corresponds to the bound 
(\ref{bound}) with the equal sign and therefore proves that 
our scheme is an \emph{optimal} explicit unitary 
realization of a quantum repeater for qubits.
Notice that we have set $\phi_2=0$, {\em i.e.} this result 
has been obtained using only one of the two probe' degrees 
of freedom.
%%%%%%%%%%%%%%%%%%%%%%%%%%%%%%
\subsection{A more general scheme}\label{ss:sm}
We now explore the possibility of generalizing our scheme for the
measurement of the spin in a generic direction, {\em i.e} for the
measurement of the observable $\sigma_m = {\mathbf R}_m^{\dag}\: \sigma_z 
\: {\mathbf R}_m$. Let us consider a scheme similar to that in Fig. 
\ref{f:sch2} with the ${\mathbf C}$ gate replaced by the gate 
${\mathbf W}=({\mathbbm I} \otimes {\mathbf R}_m)\: {\mathbf C}$, 
the corresponding map operators are given by
\begin{equation}
A'_k = {}_{mp}\langle k|\:({\mathbbm I}\otimes {\mathbf R}_m)\: 
{\mathbf C}\:|\omega\rangle_p = {}_p\langle k|{\mathbf C}|
\omega\rangle_p = A_k  \label{s1}\:,
\end{equation}
with $|k\rangle_m$ eigenstates of $\sigma_m$, 
whereas the POVM is obtained as 
\begin{eqnarray}
\Pi'_k &=& \hbox{Tr}_p\left[{\mathbf W} \: {\mathbbm I}\otimes 
|\omega\rangle_p{}_p\langle \omega| \: {\mathbf W}^{\dag}\: {\mathbbm I}
\otimes |k\rangle_{mp}{}_{mp}\langle k| \: \right]  \nonumber \\
&=& \hbox{Tr}_p\left[{\mathbf C}\: {\mathbbm I} \otimes |\omega
\rangle_p{}_p\langle\omega | \: {\mathbf C}^{\dag}\: {\mathbbm I} \otimes 
|k\rangle_p{}_p\langle k| \: \right] = \Pi_k \:.  \label{s2}
\end{eqnarray} 
The primed operators in Eqs. (\ref{s1}) and (\ref{s2}) are equal
to the operators in Eqs. (\ref{A_k}) and (\ref{Pi_k}).  
As a consequence the probabilities $p_k$, and the
conditional states $|\psi_k\rangle$ are the same as in the scheme
of the previous section.  At this point we note that the simple
inference rule $k \rightarrow |k\rangle_m$ cannot be used. In this case, in fact,
we would obtain the same fidelity  $F$ as in the previous
section, but a different fidelity $G$, and thus our repeater
couldn't be optimal.  However, it is straightforward to
re-establish the same expression of $G$ using the inference rule
$k \rightarrow |k\rangle $ (with $|k\rangle$ eigenstates of
$\sigma_z$ ). Using this procedure the fidelities become equal to
Eqs. (\ref{F_opt}) and (\ref{G_opt}) and the repeater is again
optimal.
%%%%%%%%%%%%%%%%%%%%%%%%%%%%%%%%
\subsection{Optimal quantum repeaters for qudits}\label{ss:D}
\begin{figure}[h]
\begin{center}
\includegraphics[width=0.4\textwidth]{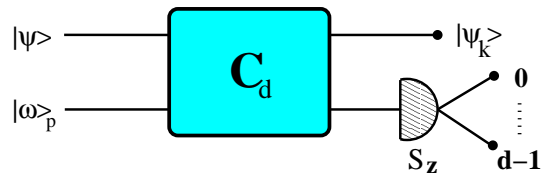}
\end{center}
\caption{Minimal implementation of optimal quantum repeater for qudits.} 
\label{f:schd}
\end{figure}

The optimal repeater for qubits described in the previous 
Sections can be generalized to obtain an optimal repeater 
for qudits. The scheme is depicted in Fig. \ref{f:schd} and 
is similar to that of Fig. \ref{f:sch2} with the 
$C_{\rm not}$ replaced by its $d$-dimensional counterpart, 
{\em i.e.} by the gate acting as ${\mathbf C}_d|i\rangle
|s\rangle_p = |i\rangle |i\oplus s\rangle_p$ 
where $\oplus$ denotes sum modulo $d$ \cite{delg}. 
The corresponding matrix elements
reads as follows ${}_p\langle s | \langle i | {\mathbf C}_d | j\rangle |
s'\rangle_p = \delta_{ij}\delta_{s,s'\oplus j}$.
%%%%%%%%%%%%%%%%
%%%%%%%%%%%%%%%%
\par\noindent
The probe qudit is prepared in the state
\begin{eqnarray}
|\omega\rangle_p = \cos\theta_2 |0\rangle_p + \gamma \sin\theta_2 \: 
\frac{1}{\sqrt{d}} \sum_{s=0}^{d-1} | s\rangle_p
\label{omd}\;,
\end{eqnarray}
where 
\begin{eqnarray}
\gamma = \frac{\sqrt{1+d \tan^2\theta_2}-1}{\sqrt{d} 
\tan\theta_2}
\label{gamma}\;,
\end{eqnarray}
is a normalization factor. As for qubits, an optimal repeater can be 
obtained exploiting a single probe' degree of freedom. 
After the interaction the spin of the probe is measured in a given
direction. In the following, having in mind the equivalence already 
shown for qubits, we refer to a scheme where the spin is 
measured in the $z$-direction. \par
The measurement operators are given by
\begin{eqnarray}
A_k = {}_p \langle k | {\mathbf C}_d | \omega \rangle_p = \sum_{ij} 
(A_k)_{ij} |i\rangle\langle j|
\label{akd}\;,
\end{eqnarray}
where 
\begin{eqnarray}
(A_k)_{ij} = \delta_{ij} \left[ \delta_{kj} \cos\theta_2 + \gamma \sin\theta_2 
\frac{1}{\sqrt{d}}\sum_s \delta_{k,j\oplus s}\right]
\label{akd_ij}\;.
\end{eqnarray}
The fidelities are evaluated using Eqs. (\ref{fidse}) 
and (\ref{akd_ij}), arriving at
\begin{eqnarray}
F&=& \frac{1}{d+1} \left[1 + \left(\cos\theta_2 + \gamma \sqrt{d} 
\sin\theta_2 \right)^2\right] \nonumber \\
G&=& \frac{1}{d+1} \left[1 + \left(\cos\theta_2 + \frac{\gamma}{\sqrt{d}} 
\sin\theta_2 \right)^2\right]
\label{Dfids}\;,
\end{eqnarray}
which may be tuned by varying the preparation of the probe {\em i.e.}
the value of $\theta_2$.
Inserting (\ref{Dfids}) into (\ref{Dbound}) we found that the bound
is saturated for $\gamma$ given by (\ref{gamma}). In other words, 
the scheme of Fig. \ref{f:schd} with a $d$-dimensional 
$C_{\rm not}$ and a probe qudit given by (\ref{omd}) provides an 
optimal quantum repeater for qudits.
%%%%%%%%%%%%%%%%%%%%%%%%%%%%%%%%
\section{Special classes of qubits}\label{s:sig}
The bound in Eq. (\ref{bound}) has been derived with the assumption 
that the incoming signal is chosen at random on the whole Bloch 
sphere. In this section we analyze whether a different choice of the 
alphabet may be used to beat the bound and, in turn, to improve the
information-disturbance trade-off. As we will see, this is indeed the 
case assuming that the input signal is chosen within a discrete set
of states, whereas a continuous subset of the Bloch sphere leads
to degraded performances. 
\par
\begin{figure}[h!]
\centering
{\includegraphics[width=0.20\textwidth ]{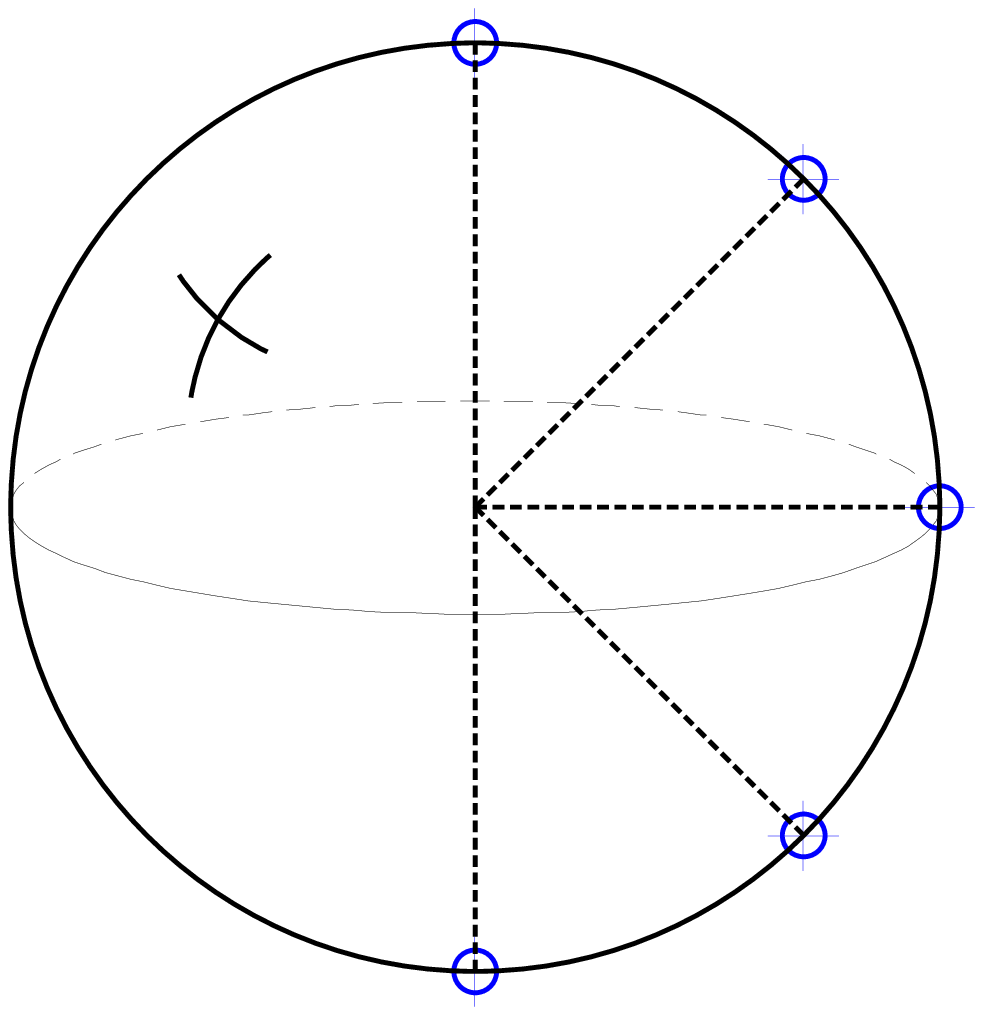}} $\;$
{\includegraphics[width=0.191\textwidth ]{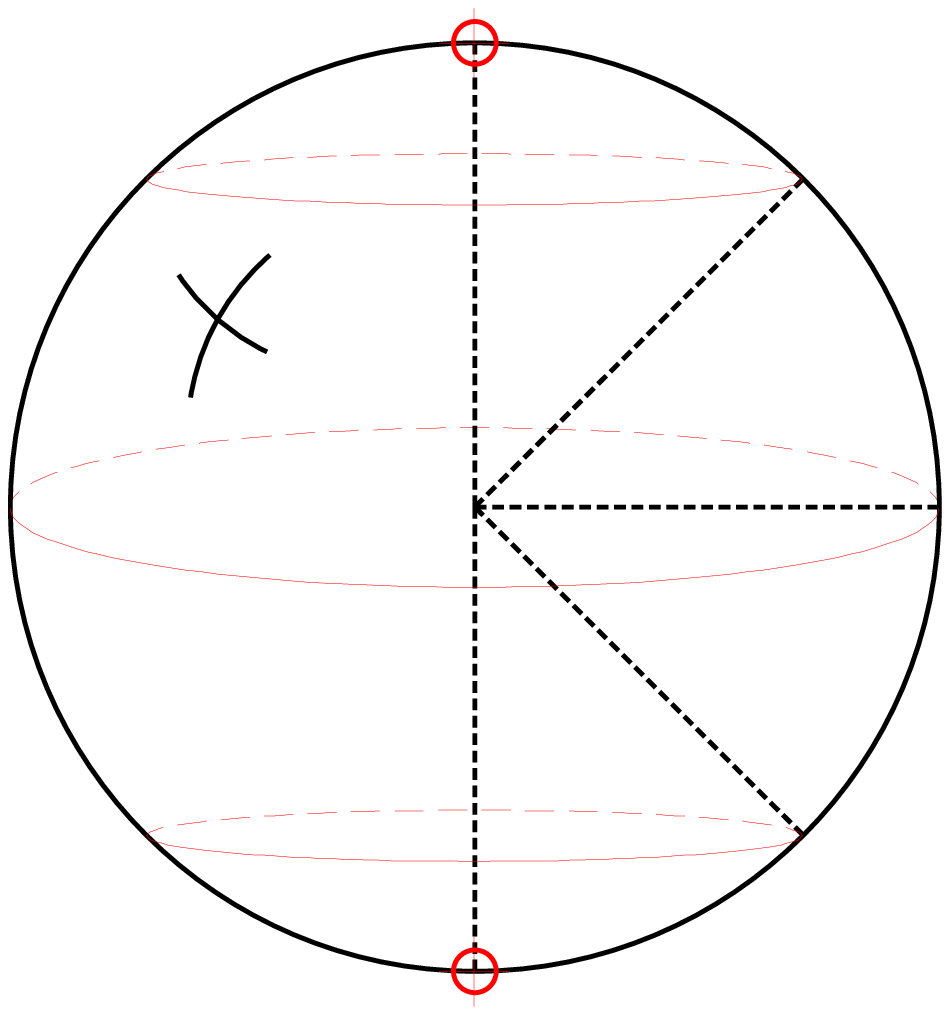}}
\caption{Left: qubits belonging to discrete class {\bf A} ($N=5$).
Right: qubits belonging to continuous class {\bf B} ($N=5$) 
\label{Classes}}
\end{figure}

Let us consider the optimal repeater of Section \ref{s:sz} 
with the input signal chosen within the following two classes 
of states
\begin{itemize}
\item[{\bf A.}] A discrete set made of $N$ states $\psi_j$ equally 
spaced in $\theta$ and with random phase $\phi$. Since the fidelities 
$F_{\psi}$ are phase-independent we set, without loss
of generality, $\phi=0$
$$ |\psi_j\rangle_A = \cos\frac{\theta_j}{2}|0\rangle + 
\sin\frac{\theta_j}{2}|1\rangle \qquad j=0,\dots,(N-1) $$
where $\theta_j=\frac{j\pi}{N-1}$.
\item[{\bf B.}] A continuous set of $2\pi\times N$ states, 
equally spaced in $\theta$ and with random phases:
$$ |\psi_{j\phi}\rangle_B = \cos\frac{\theta_j}{2}|0\rangle + 
e^{i\phi}\sin\frac{\theta_j}{2}|1\rangle \qquad j=0,\dots,(N-1) $$
where $\theta_j=\frac{j\pi}{N-1}$ and $\phi \in [0,2\pi]$.
\end{itemize}
The two sets are schematically depicted in Fig. \ref{Classes}.
%%%%%%%%%%%%%%%%%%%%%%%%%
%%%%%%%%%%%%%%%%%%%%%%%%%
\par
\begin{figure}[h] 
\includegraphics[width=0.35\textwidth]{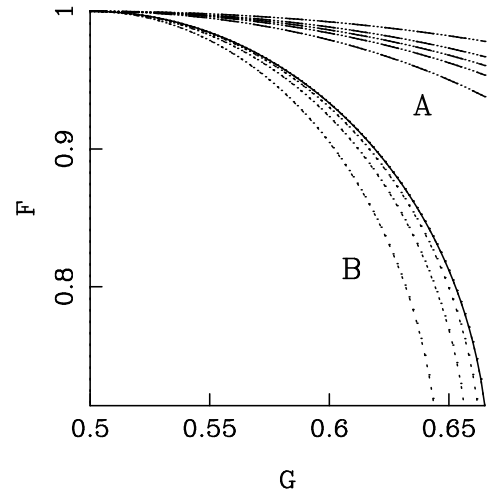} 
\caption{The functions $F(G)$ for signals of class {\bf A} 
and {\bf B} for different values of $N$. The solid line 
denotes the bound $F(G)$ imposed by Eq. (\ref{bound}). 
Dot-dashed lines denote $F_{AN} (G_{AN})$ 
whereas dotted lines are for $F_{BN} (G_{BN})$. 
We plot curves for $N=4,5,7,11$ and $N=1000$.
$F_{AN}(G_{AN})$ is always above the bound (\ref{bound})
and decreases with increasing $N$. 
$F_{BN}(G_{BN})$ is always below the bound (\ref{bound})
and increases with increasing $N$. 
} \label{PlotFs}
\end{figure}

\noindent
Using  Eqs. (\ref{F_psi}) and (\ref{G_psi}) we find that fidelities 
corresponding to states $|\psi_j\rangle_A$ and $|\psi_{j\phi}\rangle_B$ 
(with the same $\theta_j$ and different phases) are given by
\begin{eqnarray}\label{fj}
F_j &=& \cos^4\frac{\theta_j}{2} + \sin^4\frac{\theta_j}{2} + 
4\sin^2\frac{\theta_j}{2}\cos^2\frac{\theta_j}{2}\sin\frac{\theta_2}{2}
\cos\frac{\theta_2}{2} \nonumber \\ 
&=& \frac12 \left[(1+\cos^2\theta_j) + 
\sin\theta_2 (1-\cos^2\theta_j)\right] \\
G_j &=& (\cos^4\frac{\theta_j}{2} +\sin^4\frac{\theta_j}{2}) 
\cos^2\frac{\theta_2}{2} + 2\sin^2 
\frac{\theta_j}{2}\cos^2\frac{\theta_j}{2}\sin^2\frac{\theta_2}{2}
\: \nonumber \\
&=& \frac12 \left(1+ \cos^2\theta_j \cos \theta_2\right) 
\label{gj}
\end{eqnarray}
The mean fidelities for class {\bf A} are given by 
\begin{eqnarray}
F_{AN}&=& \frac{1}{N}\sum_{j=0}^{N-1}F_j = 
\frac{1 + 3N +(N-1)\sin\theta_2}{4N} \nonumber \\
G_{AN}&=& \frac{1}{N}\sum_{j=0}^{N-1}G_j  = 
\frac{1 + 3N + (N-1)\cos\theta_2}{4N}\:,
\end{eqnarray}
from which we obtain
\begin{widetext}
\begin{eqnarray}
F_{AN}(G_{AN}) = \frac{1}{4N}\left(1 + 3N +\frac{N-1}{N+1}
\sqrt{(N+1)^2 - 4 N^2 (1-2 G_{AN})^2} \right)
\:.\end{eqnarray}
$F_{AN}$ is a monotonously decreasing function of $N$ 
and, as it can be seen in Fig. \ref{PlotFs}, 
is above the bound set by Eq. (\ref{bound}) for any value 
of $N$. Therefore, by transmitting a discrete alphabet of
symbols, we can beat the bound (\ref{bound}), {\em i.e.} 
the protocol is more convenient than the 
transmission of the whole class of qubits.
\par\noindent
As concerns class {\bf B}, the mean fidelities are evaluated 
as follows
\begin{eqnarray}
F_{BN}= \frac{1}{2\pi\sum_{j=0}^{N-1}\sin\theta_j}
\sum_{j=0}^{N-1}2\pi\sin\theta_j F_j \qquad
G_{BN} \frac{1}{2\pi\sum_{j=0}^{N-1}\sin\theta_j}
\sum_{j=0}^{N-1}2\pi\sin\theta_j G_j 
\end{eqnarray}
For even $N$ we obtain
\begin{eqnarray}
F_{BN} &=& \frac14 \left[3 + \sin\theta_2 + 2ie^{\frac{i(3N-1)\pi}{2(N-1)}}
(1 + \sin\theta_2) - ie^{\frac{i(5N-1)\pi}{2(N-1)}}(3 + \sin\theta_2) \right]
\nonumber \\
G_{BN} &=& \frac14 \left[2 + \cos\theta_2 + 2ie^{\frac{i(3N-1)\pi}{2(N-1)}} -
ie^{\frac{i(5N-1)\pi}{2(N-1)}}(2 + \cos\theta_2)\right] \label{x1}
\:,
\end{eqnarray}
\end{widetext}
whereas for odd $N$
\begin{eqnarray}
F_{BN} &=& \frac{1 + \sin\theta_2 + \cos\frac{\pi}{N-1}(3+
\sin\theta_2)}{2(1 + 2\cos\frac{\pi}{N-1})} \nonumber \\
G_{BN} &=& \frac{1 + \cos\frac{\pi}{N-1}(2+\cos\theta_2)}{2(1 
+ 2\cos\frac{\pi}{N-1})} \label{x2} 
\end{eqnarray}
Using Eqs. (\ref{x1}) and (\ref{x2}) we have calculated 
the explicit function $F_{BN}(G_{BN})$. The resulting expression
is quite cumbersome and will not be reported here. In 
Fig. \ref{PlotFs} we show the function $F_{BN}(G_{BN})$ 
for different values of $N$. All the curves are below the bound 
curve (\ref{bound}), approaching it for $N\rightarrow \infty$.
Therefore, if we need to transmit a continuous alphabet, it's more
effective to transmit qubits on the whole Bloch sphere rather 
than on a continuous subset.
%%%%%%%%%%%%%%%%%%%
%%%%%%%%%%%%%%%%%%%
\par\noindent
A general condition may be found for an alphabet of signals to beat
the bound (\ref{bound}). For an unspecified class of states the 
fidelities may be evaluated using Eqs. (\ref{fj}) and (\ref{gj}). 
We have  that 
\begin{eqnarray}
F&=& \frac12 [1+\overline{\cos^2\theta}+ \sin\theta_2 (1-\overline{\cos^2\theta})]
\nonumber \\ 
G &=& \frac12 (1+\cos\theta_2 \overline{\cos^2\theta})
\label{fgun}\;,
\end{eqnarray}
where $\overline{(\ldots)}$ denotes the average over the alphabet. 
Substituting Eqs. (\ref{fgun}) in Eq. (\ref{bound}) we found that 
any class of states violating the following inequality
\begin{align}
\Big[
\overline{\cos^2\theta}-\frac13 + 
(1-\overline{\cos^2\theta}) &    \sin\theta_2
\Big]^2 \nonumber \\ &+ 4\cos^2\theta_2\: 
\overline{\cos^2\theta}^2 \leq \frac49
\end{align}
provides a better information-disturbance trade-off than randomly
distributed signals.
As an example, for $\theta_2=0$ {\em i.e.} for the maximum value 
of the estimation fidelity $G$ the bound (\ref{bound}) 
is surpassed for classes of states for which $\overline{\cos^2\theta} > 1/3$.
%%%%%%%%%%%%%%%%%%%
\section{Conclusions}\label{s:outro}
In this paper unitary realizations of quantum repeaters, {\em
i.e.} nondemolitive estimations of qubits and qudits, have been
suggested. The schemes are {\em minimal}, {\em i.e} involve a
single probe system in addition to the signal, and {\em optimal
i.e.} they obtain the maximum information with the minimum amount
of noise allowed by quantum mechanics for randomly distributed
signals. \par We then analyzed the performances of optimal
repeaters on different classes of signals, corresponding to
alphabets that are subsets of the whole Bloch sphere. We derive a
general condition that a class of states should satisfy to beat
the bound (\ref{bound}) showing that discrete alphabets can
beat this bound, whereas continuous alphabets lead to inferior
performances.
%%%%%%%%%%%%%%%%%%%%%%%%%%%%%%%%%%%%%%%%%%%%%%%%%%%%%%%%%%%%%%%%

%%%%%%%%%%%%%%%%%%%%%%%%%%%%%%%%%%%%%%%%%%%%%%%%%%%%%%%%%%%%%%%%
\end{document}